\documentstyle[11pt,newpasp,twoside,epsf]{article} \def\edcomment#1{\iffalse\marginpar{\raggedright\sl#1\/}\else\relax\fi}
\reversemarginpar

\begin{document} \title{Evidence for Galaxy Formation at High Redshift}

\author{Tom Shanks, Nigel Metcalfe, Dick Fong} \affil{Physics Department, Durham University, Science
Laboratories, South Road, Durham DH1 3LE, UK} 

\author{Henry McCracken} \affil{LAS, Traverse du Siphon, Les Trois Lucs, F-13102 Marseille, France} 

\author{Ana Campos}\affil{Instituto de Matematicas y Fisica Fundamental, CSIC, Spain}

\author{David Thompson} \affil{California Institute of Technology, MS 320-47, Pasadena, CA 91125,
USA}

\begin{abstract}  Metcalfe et al. (1995, 1996) have shown that galaxy
counts from the UV to the near-IR are well-fitted by simple evolutionary models 
where the space density of galaxies  remains constant with look-back time  while
the star-formation rate rises exponentially. We now extend these results, first by using
data from the Herschel Deep Field to show that these same models give detailed
fits to the faint galaxy $r-i:b-r$ colour-colour diagram. We then use these
models to predict the number counts of high redshift galaxies detected by the
Lyman break technique. At $z\approx3$ there is almost exact agreement between our
prediction and the data, suggesting that the space density of galaxies at
$z\approx3$ may be close to its local value. At $z\approx4$ the space density of
bright galaxies remains unchanged; however, the space density of dwarf galaxies
is significantly lower than it is locally, suggesting that we have detected an
epoch of dwarf galaxy formation at $z\approx4$. Finally, significant numbers of
Lyman-break galaxy candidates are also detected at $z\approx6$ in the Hubble and
Herschel Deep Fields; taking this observation together with a number of recent
detections of spectroscopically confirmed $z\approx6$ galaxies suggests that the
space density of bright galaxies at  $z\approx6$ remains comparable to the
local space density and thus that the epoch of formation of bright galaxies may
lie at yet higher redshift.
\end{abstract}

\section{Introduction} The epoch of galaxy formation has been the subject of much
recent controversy.  Initially, observational evidence for a late epoch of
galaxy formation appeared to come from analysis of the Canada-France Redshift
Survey and the Hubble Deep Field where the universal star-formation rate (SFR)
appeared to peak at $z\approx1$ (Lilly et al. 1995, Madau et al. 1996). However,
it was also suggested that the CFRS/HDF data could still be consistent with
models (Bruzual \& Charlot 1993) with a constant space density of galaxies whose 
SFR rises exponentially to redshifts $z\ga2$ (Metcalfe et al. 1996, Shanks et
al. 1998). Indeed, if these `pure luminosity evolution' models are modified by
assuming a small amount of dust internal to spirals and  a dwarf-dominated IMF
for early-type galaxies, then   they  appeared to give very good fits to galaxy
number count data from the UV to the near-infrared, at least in the case of low
$\Omega_0$ (Metcalfe et al. 1996). The recent detection of high redshift galaxies
as sub-mm sources (Smail et al. 1997, Hughes et al. 1998, Barger et al. 1998)
also appears to support the idea that the SFR may rise to much higher redshifts.
The `pure luminosity evolution' models with internal spiral dust absorption have
also been shown to be capable of producing fits to the sub-mm galaxy counts and
the optical/near-IR galaxy counts simultaneously (Busswell \& Shanks 2000).
Details of further constraints from the UV colours of galaxies on the dust
extinction law  and its evolution with redshift are given by Metcalfe et al.
(2000).

In the Einstein-de Sitter Universe, with its smaller volume, it has been
known for some time that such simple models only fit the optical counts to
B$\la$25mag (Yoshii \& Takahara 1988, Guiderdoni \& Rocca-Volmerange 1990,
Koo et al. 1993) before underestimating the counts at fainter magnitudes. In
this case, a new population of faint galaxies must be postulated; these extra
galaxies could be the result of a process of dynamical merging at high
redshift (Kauffmann \& Charlot 1998). Alternatively, as we assume here, they
could be due to a population of `disappearing dwarfs', denoted `dE', that are
luminous at high redshift but then dim dramatically to be virtually
undetectable at low redshift (Babul \& Rees 1992, Metcalfe et al. 1996, 2000)

\begin{figure} \plotone{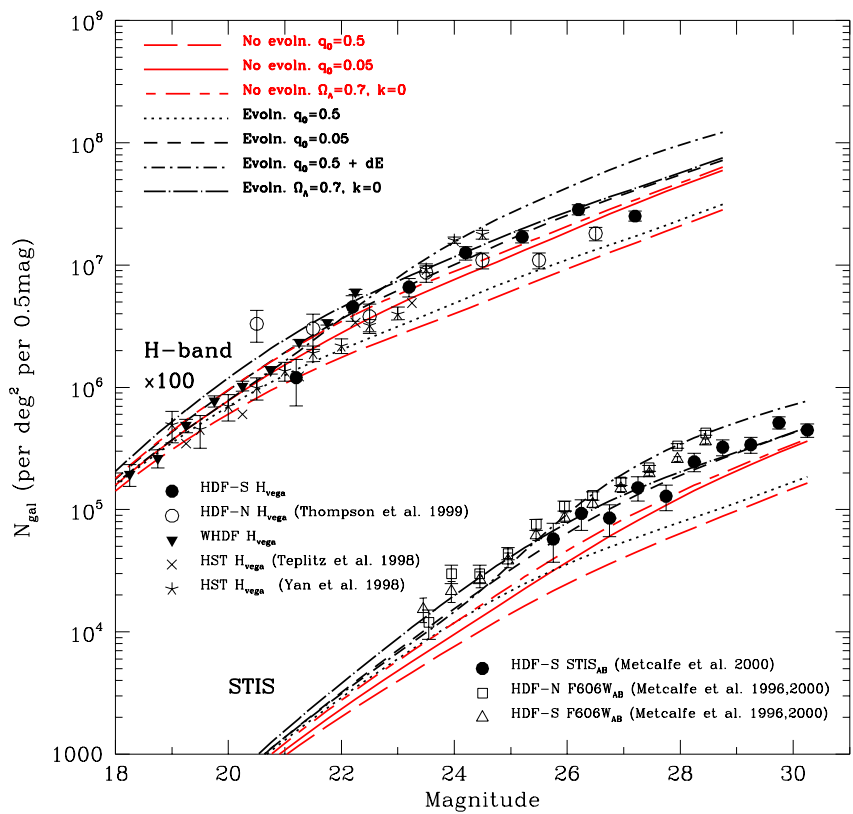}
\caption{ Lower points; HST STIS Hubble Deep Field South galaxy counts in the Clear
passband from the data of Williams et al. (2000), analysed by techniques described
elsewhere (Metcalfe et al. 2000). Also shown are the HST WFPC-2 R(F606W) counts in
the HDF-N,-S fields; the STIS counts are slightly lower than, but still
statistically consistent with, these data in the range of overlap. Upper points;
HST NICMOS HDF-N,-S galaxy count data in the infra-red H (F160W) passband from our
analysis (Metcalfe et al. 2000) of the data of Williams et al. (2000) in the case
of HDF-S and directly from Thompson et al. (1999) in the case of HDF-N.  The counts
have been scaled upwards by a factor of 100 for clarity. HST NICMOS data from other
authors are also shown (Teplitz et al. 1998, Yan et al. 1998). The WHDF data comes
from a 20hr exposure at the Calar Alto 3.5m telescope using the $\Omega'$ camera.
The field-of-view is 7$'\times$7$'$ and the galaxy counts reach H=22.5. The models
shown are those of Metcalfe et al. (1996) applied to both passbands (see text).}
\end{figure}

\section{Galaxy Counts and Colours} We now test how well these models fit new
ultra-deep optical and IR count data from HDF-N (Williams et al. 1996, Thompson et
al. 1999), HDF-S (Gardner et al. 2000, Williams et al. 2000) and the William
Herschel Deep Field (WHDF). Full details of our analyses of the Hubble and Herschel
Deep Fields are given elsewhere (Metcalfe et al. 1995, 1996, 2000). The new optical
($R_{AB}$) counts from HST STIS observations in the HDF-S are shown  in the lower
part of Fig. 1. The STIS passband is centred at 5850\AA\, and has a width of
$\sim$4000\AA. The field-of-view is 50$\times$50 arcsec$^2$. The exposure time was
43hrs and these are the deepest optical galaxy counts, reaching beyond R$_{AB}$=30.
Also shown are the HDF-N,-S WFPC-2 R counts.  Above these in Fig. 1 are the new
H-band counts from  NICMOS observations in HDF-N,-S. The field-of-view was
$\sim$50$\times$50 arcsec$^2$. The exposure time was 36.5hrs for HDF-N and 48.3hrs
for HDF-S and these are the deepest galaxy counts in the near-IR so far produced. 
Fig. 1 also shows the WHDF H counts from Calar Alto 3.5-m observations plus H counts
from other authors. The models shown are those  of Metcalfe et al. (1996, 2000)
applied to both passbands. The no-evolution models generally underestimate the
optical data. The q$_0$=0.05 and $\Omega_\Lambda$=0.7 evolutionary models give
excellent fits to the faintest limit of the HST STIS R counts. The q$_0$=0.5 $+$ dE
dwarfs model includes the `disappearing dwarf' galaxy population previously invoked
to improve the galaxy count fit of the q$_0$=0.5 evolutionary model (Metcalfe et
al. 1996) and this model also remains consistent with the new, fainter data. This
model slightly overestimates the faintest counts at $R_{AB}\ga29$ and $H\ga25$,
possibly indicating a slightly smaller contribution from the dE population than
assumed here (Metcalfe et al., 1996). Nevertheless, we also regard this model as
remaining in reasonable agreement with the ultra-deep galaxy counts data. At H,
where the non-evolving models lie closer to the data than in the optical, all the
evolving models except  the q$_0$=0.5 models without the dE component, provide a
reasonable fit to the counts to the faintest $H\sim28$ limit of the HST NICMOS
data. We conclude that the predictions  for the evolutionary models continue to
agree with the data to these new, very deep  completeness limits of $R_{AB}$=30 and
$H=28$.

\begin{figure}
\plotfiddle{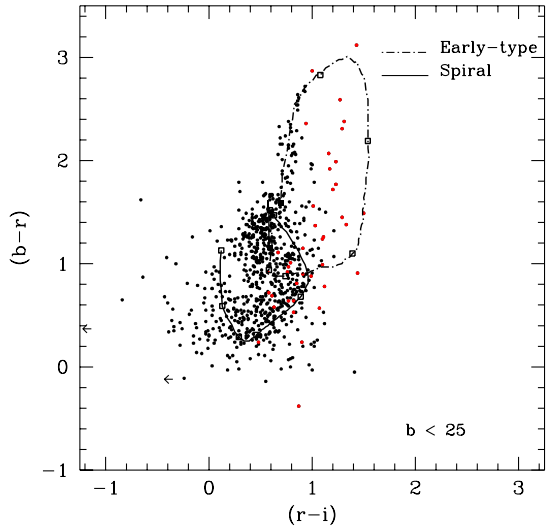}{3in}{0}{40}{40}{-120}{-50}
\caption{The b-limited $r-i:b-r$ colour-colour diagram from the
Herschel Deep Field data, showing the evolutionary tracks predicted  from our
Bruzual \& Charlot models for essentially two galaxy types for $0\la z\la 3$. Boxes
indicate 0.5z intervals. A plume of early-type galaxies can be seen which rise
from $b-r\approx1.5, r-i\approx0.7$ to redder $b-r$ colours before turning
bluewards again, as increasing star-formation at $z\ga 0.5$ begins to affect the
redshifted UV light. The spirals follow  a horseshoe-shaped track below the
early-types. This result clearly shows that there are just two main forms for the
evolutionary behaviour of faint galaxies and the excellent fit of the Bruzual and
Charlot models suggests that the main distinguishing feature is the star-formation rate,
with $\tau$=2.5Gyr assumed here for the early-types and $\tau$=9Gyr for the spirals.
} \end{figure}

Beyond $R_{AB}\approx26$ and $H\approx23$ we are looking down the luminosity
function of galaxies and seeing intrinsically faint galaxies(M$_H\sim$-12),
which on consideration of their colours are mostly at $z\approx2$ (Metcalfe
et al. 1996, 2000). Thus intrinsically faint galaxies at relatively low
redshifts dominate the galaxy number counts at the faintest limits. In the
low q$_0$ case, the excellent fit of the evolutionary models suggests that
the slope of the faint galaxy luminosity function at $z\approx$2 may be
consistent with that measured locally. The need for the extra dwarf
population to fit the optical counts in the $q_0$=0.5 case suggests that
here the luminosity function slope must steepen by $z\approx2$ to match the
faint count data.

Finally, we compare the model  predictions for the  $r-i:b-r$ colour-colour
diagrams with  the data from the WHDF. Fig. 2 shows how the predictions show a
distinctive plume of early-type galaxies which first rise from $b-r\approx1.5,
r-i\approx0.7$ to redder $b-r$ colours before turning bluewards again, as the
increasing SFR at high redshift begins to affect the redshifted UV light. In the
case of the spirals a horseshoe-shaped curve is predicted reaching out to
$z\approx3$ at its faint limit. The agreement between models and data in both
cases is quite remarkable, suggesting that for all their simplicity these models
share the essential qualities required by the data; for example, the faint
galaxies seem to divide into just two evolutionary classes corresponding to early
and late types as we assume. And the parameter differentiating the
evolution in each class appears to be the star-formation rate, with an
exponential e-folding time of $\tau=2.5$Gyr for early-types and $\tau=9$Gyr for
spirals.

\begin{figure}
\plotfiddle{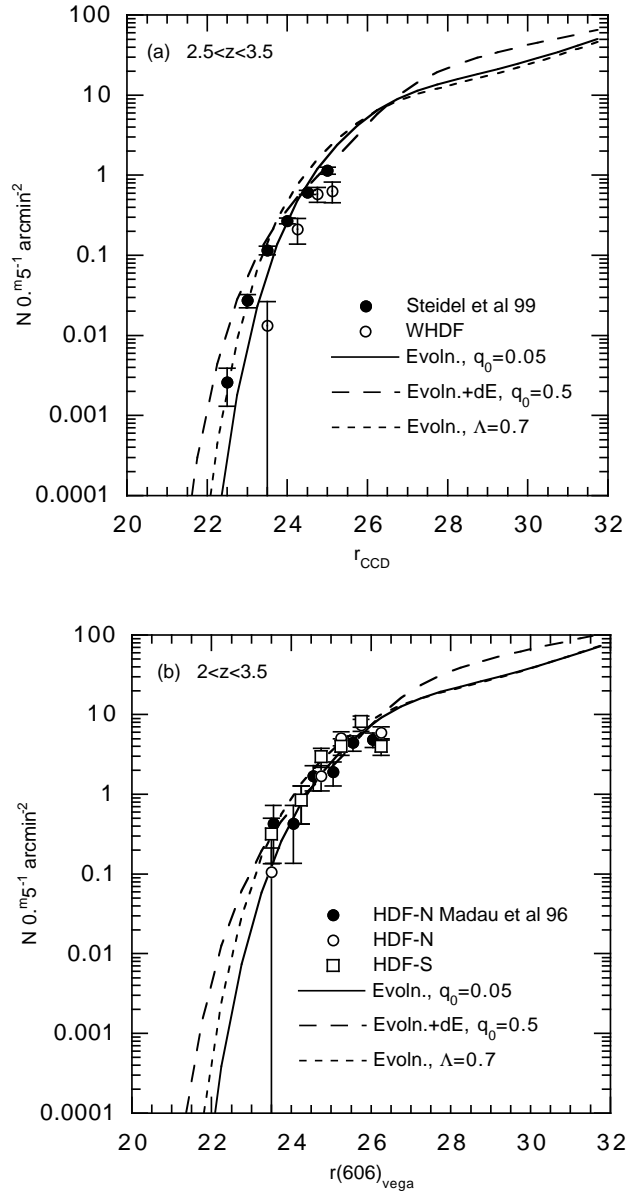}{7in}{0}{65}{65}{-185}{-10}
\caption{(a) The $r_{CCD}$ number count of $2.5\la z\la 3.5$
ground-based UV dropout galaxies compared to our evolutionary model
predictions.  In the q$_0$=0.5 $+$ dE case the contribution of the dE
population can be seen at $r>25$. The WHDF data reasonably agrees with
the data of Steidel et al. (1999) and overall we regard the agreement
between the three evolutionary models and the data as
excellent. (b) The r (F606W$_{vega}$) number count of $2\la z\la3.5$
HDF UV dropout galaxies compared to the same models as in (a). Again
the agreement between these fainter data and the three 
evolutionary models is excellent.} \end{figure}

\begin{figure}
\plotfiddle{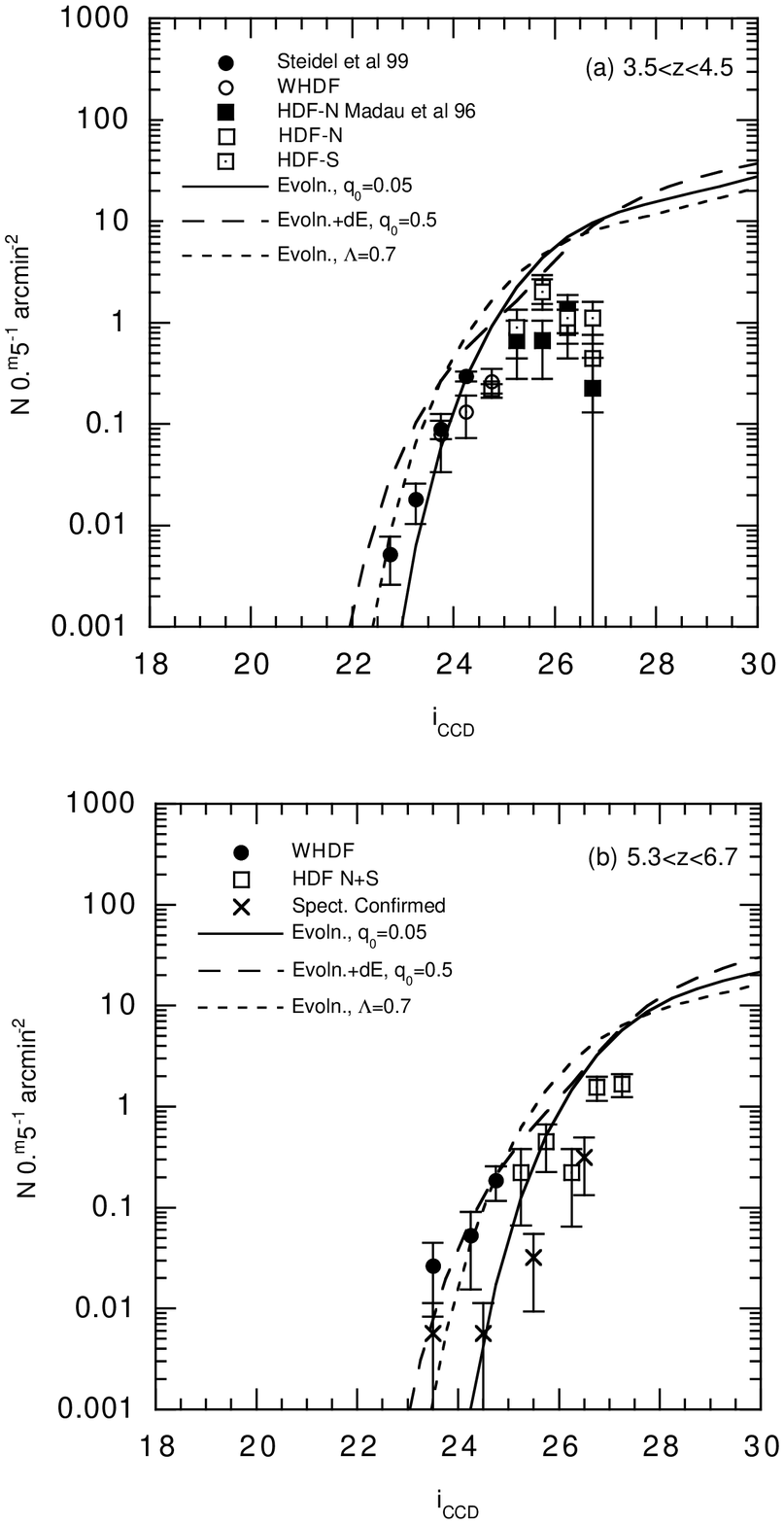}{7in}{0}{65}{65}{-185}{-10}
\caption{(a) The $i_{CCD}$ number count of $3.5\la z\la4.5$ B dropout
galaxies from the WHDF and HDF data, compared to our model
predictions. The models fit well at bright magnitudes but are
5$\times$ too high at faint magnitudes, indicating that luminous
galaxies may form at $z\ga4$ while dwarves form at $z\la4$.  (b) The
$i_{CCD}$ number count of $5.3\la z\la 6.7$ R dropout galaxies from
the WHDF and HDF data (Metcalfe et al. 2000). The crosses are
based on 6 spectroscopically confirmed $z\approx6$ galaxies.  The
models show good agreement for the brighter galaxies, indicating that
luminous galaxies form at $z\ga6$. }
\end{figure}

\section{`Drop-out' galaxies} We now use these models to predict the number
counts of galaxies detected by the Lyman break technique (Steidel et al.
1999) at $2\la z\la 7$. We first consider the $z\approx3$ range, where the
data comes from detecting the Lyman break galaxies via the UV dropout
technique in ground-based data (Steidel et al. 1999, Metcalfe et al. 2000)
and at fainter limits from the Hubble Deep Fields N,S (Madau et al. 1996,
Metcalfe et al. 2000). In Fig. 3a, we then simply compare the observed
ground-based number counts of the $z\approx3$ galaxies with the model
predictions, with the same comparison for the HST-based data being shown in
Fig. 3b. We see that in both cases at $z\approx3$ there is excellent
agreement between the three basic evolutionary models and the data. Thus
models which were last adjusted to fit the Keck B$<$24 n(z) at $z\approx1$
are now found to give a strikingly good fit to what is effectively the galaxy
luminosity function at $z\approx3$. Since it is the spirals which dominate
the bright part of the model luminosity functions at z=3, this means that the
Lyman break galaxies might be identified with the bright end of the evolved
spiral luminosity function. This result is consistent with the flat SFR:z
relation between $1\la z \la3$ found by Steidel et al. (1999) and the slightly
rising SFR:z relation predicted by the models in this range (Metcalfe et al.
2000); the models predict only a small amount of evolutionary brightening in
the range $1\la z \la3$ and the constant space density of spiral galaxies then
ensures the SFR:z relation remains fairly flat. In the fainter, HST-based
data in Fig. 3b, the models continue to fit well. However, it should be noted
that the q$_0$=0.5 model relies on the contribution of the `disappearing
dwarfs' to maintain agreement with the faint count data; it can be seen that,
without this dE contribution, the q$_0$=0.5 model would start to
underestimate the counts.

In Fig. 4a we now compare the number count of $z\approx4$ Lyman break
galaxies detected by B dropout to the same models. At bright
magnitudes, where the data comes mainly from Steidel et al.
(1999), the models appear to fit the data well, suggesting that the
bright spirals may have been in place even as early as
$z\approx4$. However, at fainter magnitudes where the data comes from
the HDF-N,-S and the WHDF, the simple pure luminosity evolution models
at last appear to be breaking down, with the models generally
overestimating the $z\approx4$ faint galaxy luminosity function.  The
observed faint ($i_{CCD}>24.5$) galaxy count falls below our
predictions by a factor of $\approx$5, indicating that while bright
L$^*$ galaxies may have been in place at $z\approx4$, the faint galaxy
population may have formed in the range $3\la z\la 4$.  Steidel et
al. (1999) also noted that the galaxy luminosity function at
$z\approx4$ appeared flatter than at $z\approx3$ but speculated that
this may be due to the HDF-N being an unrepresentative field. The fact
that the same result is seen in the WHDF and HDF-S argues that this
result may be real. The result also explains why the SFR density of
Steidel et al. (1999) appears to be constant between $3\la z\la 4$
when measured from bright ground-based data, but appears to fall at
$z\approx4$ when measured from the fainter HDF data.

In Fig. 4b we extend this approach, now comparing the number counts of
$z\approx6$ Lyman break galaxies detected by R dropout in the HDF and
WHDF to the same models. We have taken galaxies with $r-i>1.5$ and
$i-H<2.2$ in the WHDF and (F606-F814)$_{vega}>$2.44 in the HDF-N,-S to
define consistently the redshift range 5.3$\la$z$\la$6.7 or
$z\approx6$ as suggested by the models.  Again we see that the
observed number density of bright WHDF galaxies is well fitted by the
models. The fainter galaxies from the HDF remain a factor of
$\approx$5 lower than predicted. Since the HDF data probes less deeply
into the galaxy luminosity function than at $z\approx4$, this is
consistent with a further decrease in the number of faint galaxies
relative to the local density.  The crosses in Fig. 4b show the number
counts we have compiled from 6 galaxies with Keck or HST spectroscopy
(Dey et al. 1998, Weymann et al.  1998, Spinrad et al. 1998, Chen et
al. 1998). It can be seen that although there is reasonable agreement
at the faint end with our unconfirmed HDF-N,-S R dropout candidates,
at the bright end there are somewhat fewer galaxies detected.  This
disagreement may be partly due to the difficulty of detecting Ly
$\alpha$ emission against an increasingly bright sky at these longer
wavelengths and partly due to contamination of our WHDF candidates by
lower redshift early-type galaxies.  However, the fact that a
population of spectroscopically confirmed $z\approx6$ galaxies have
already been identified and that even larger numbers of R dropout
candidates exist, again suggests that significant numbers of bright
galaxies had formed by $z\approx6$, pushing the epoch of formation of
giant galaxies back to even higher redshifts.

\section{Conclusions} Thus the picture of galaxy formation which is suggested
is that bright galaxies form first at $z\ga4$ and then faint galaxies form
later at $3\la z\la4$. This is similar to the `downsizing' scenario
suggested in other data at $z\approx1$ (Cowie et al. 1997) and may possibly
be connected to claims of evolving, faint luminosity function slopes at even
lower redshifts (Brinchmann et al. 1998).  This scenario is unexpected in
hierarchical galaxy formation theories where dwarf galaxies are expected to
form {\it earlier} than giant galaxies. In a CDM model with feedback, this
order may be reversed with supernovae winds blowing gas out of dwarf galaxies
preferentially because of their shallower gravitational potentials. However,
suppression of star formation at $z\ga1$ is usually required in such models
to produce the observed luminosity function at the present day (Brinchmann et
al. 1998); it thus remains to be seen how well a CDM model will deal with the
large numbers of luminous galaxies detected between $4\la z\la 7$. New
observations of the galaxy luminosity function at even higher redshifts are
now required to determine the redshift of formation of the brightest
galaxies.

\section*{Acknowledgments}

We gratefully acknowledge G. Bruzual and S. Charlot for use of their
isochrone synthesis code. We also acknowledge the use of  the Hubble Space
Telescope Deep Field data. We further acknowledge use of the Calar Alto 3.5-m
telescope and the $\Omega'$ Camera. The WHT is operated on the island of La
Palma by the Isaac Newton Group at the Spanish Observatorio del Roque de los
Muchachos of the Instituto de Astrofisica de Canarias.  H.J. McCracken and N.
Metcalfe  acknowledge PPARC funding.

\end{document}